\documentclass[prb,showpacs]{revtex4}
\usepackage{graphics} 
\usepackage{epsfig} 

\begin{document}   

\title{Natural linewidth analysis of d-band photoemission from
  Ag(110)}

\author{A.\ Gerlach}
\email[E-mail: ]{agerlach@physik.uni-kassel.de}
\author{K.\ Berge} 
\email[E-mail: ]{berge@physik.uni-kassel.de}
\author{T.\ Michalke} 
\email[E-mail: ]{michalke@physik.uni-kassel.de}
\author{A.\  Goldmann} 
\email[E-mail: ]{goldmann@physik.uni-kassel.de}

\affiliation{Fachbereich Physik, Universit\"at Kassel,
  Heinrich--Plett--Str.\ 40, D--34132 Kassel, Germany}

\author{C.\ Janowitz}
\email[E-mail: ]{christoph.janowitz@physik.hu-berlin.de}
\author{R.\ M\"uller}
\email[E-mail: ]{ralph.mueller@physik.hu-berlin.de}
\affiliation{Institut f\"ur Physik, Humboldt-Universit\"at zu Berlin, 
  Invalidenstr.\ 110, D--10115 Berlin, Germany}

\date{\today}

\begin{abstract}  
  We report a high-resolution angle-resolved study of photoemission
  linewidths observed for Ag(110). A careful data analysis yields
  $k$-resolved upper limits for the inverse inelastic lifetimes of
  $d$-holes at the X-point of the bulk band structure. At the upper
  $d$-band edge the hole-lifetime is $\tau_h \geq 22\,$fs, i.e.\ more
  than one order of magnitude larger than predicted for a
  free-electron gas.  Following calculations for $d$-hole dynamics in
  Cu (I.\ Campillo et al.\/, Phys.\ Rev.\ Lett., in press) we
  interpret the lifetime enhancement by a small scattering
  cross-section of $d$- and $sp$-states below the Fermi level.  With
  increasing distance to $E_F$ the $d$-hole lifetimes get shorter
  because of the rapidly increasing density of $d$-states and
  contributions of intra-$d$-band scattering processes, but remain
  clearly above free-electron-model predictions.  
  
  \pacs{71.45.Gm, 72.15.Lh, 79.60.-i}
\end{abstract}

\maketitle

\section{Introduction} 
The dynamics of excited electrons at metal surfaces plays a key role
in understanding basic processes like chemical reactions at surfaces
and transient magnetization effects on a femtosecond time-scale. While
this time regime is now accessible by ultrashort laser pulse
pump-probe techniques, which for example allow to determine the
occupation time of the intermediate level with high precision, the
detailed interpretation of such results has generated considerable
controversy. The underlying problem is easily recognized: The observed
energy relaxation times do not represent the lifetime of a single
excited electron, but result from a convolution of several complex and
not yet well understood occupation/decay-channels like capture of
time-delayed secondary electrons and removal of electrons by ballistic
transport out of the region probed experimentally. Therefore a basic
understanding of all steps involved is required.
\\

In this context low-index surfaces of the noble metals have been
investigated very detailed both experimentally
\cite{petek97,pawlik97,knoesel98,petek00} and theoretically
\cite{echenique00,knorrenpress,knorrensubmitted}. The experimental
data for excited state electron lifetimes of copper at energies above
the Fermi energy $E_F$ show an unexpected dependence on distance from
$E_F$, on photon energy and laser pulse duration
\cite{pawlik97,knoesel98,petek00}: The intermediate state lifetime
probed by pulse-probe techniques exhibits a peak at energies just
above the threshold for excitation of $d$-holes. The dispute is
whether \cite{knoesel98} or not\cite{petek00} the delayed decay of
$d$-holes via an Auger process is responsible for the apparent
increase of lifetime within the observed peak. The recent calculations
by Knorren et al.\ \cite{knorrenpress,knorrensubmitted} are able to
model this peak quantitatively by including the contributions of
photoexcitation, electron-electron scattering, secondary electrons
generated both from scattering cascades and Auger decay of $d$-holes,
and finally transport of excited carriers out of the detection region
\cite{knorrenpress}.  Their model is based, however, on the Boltzmann
equation which uses the $d$-hole lifetime $\tau_h$ as an adjustable
parameter, with $\tau_h = 35\,$fs \cite{knorrensubmitted}.
\\

In this context an independent determination of $d$-hole lifetimes is
highly desired. Up to now only few studies have attempted to measure
such data. Lower bounds of $\tau_h = 24\,$fs \cite{petek99} and $\tau_h
= 26\,$fs \cite{matzdorf99} have been reported at the top of the Cu
$d$-bands. Therefore we have started a systematic investigation of
$d$-hole lifetimes in noble metals using natural-linewidth analysis of
high-resolution angle-resolved one-photon photoemission spectra.  Data
for several well-defined $k$-space points within the Cu bulk band
structure have been reported elsewhere \cite{gerlach10}. To check for
systematic trends we extend our studies to the isostructural and
isoelectronic surfaces of Ag. In the present paper we report new
high-resolution data for Ag(110), and discuss them in the context of
available knowledge.

\section{Experimental details} 
The experiments with synchrotron radiation were performed at the
storage ring Bessy I in Berlin. Normal-emission photoelectron spectra
were collected using the 2m-Seya beamline and a high-resolution
photoemission station described and characterized in detail in Ref.\ 
\onlinecite{janowitz99}. The sample is mounted on a manipulator
cryostat having five degrees of freedom: $x$-, $y$-,$z$-translation,
rotation around the manipulator axis and rotation around the surface
normal.  The sample temperature could be varied between $20\,$K and
room temperature. A sample load lock system with a transfer rod allows
to decouple the sample from the manipulator for preparation by argon
ion bombardment and annealing. The energy resolution including the
photon monochromator was in most cases set to $\Delta E = 27 \pm
5\,$meV as verified by the analysis of the Fermi edge emission taken
at $T = 22\,$K.  Angular resolution was $\Delta\theta = \pm
1^{\circ}$, which is sufficient for normal emission spectra taken at
flat bands, i.e.\ just at symmetry points. The Ag(110) crystal was
oriented to $\pm 0.25^{\circ}$. It was $1.5\,$mm thick and had a
diameter of about $10\,$mm.  Its high surface quality was
characterized before by experiments in our home laboratory, using a
\textsc{SCIENTA} high-resolution photoelectron spectrometer. In these
measurements we obtained very sharp LEED spot profiles and, even more
relevant, a sharp photoemission peak from the Shockley-type surface
state residing at an initial state energy $E_i = -0.055\,$eV at $T =
300\,$K \cite{gerlach99}. These observations indicate that no
significant linewidth broadening due to defect scattering is to be
expected for the data presented below.

\section{Results} 
\subsection{Measuring hole-lifetimes} 
The spectral linewidth of a photoemission peak is generally determined
by both the lifetime of the photohole and the excited state electron
lifetime. In fact the linewidth is very often dominated by the final
state contribution \cite{smith93,matzdorf98}. There are only two
situations where the hole-lifetime is reflected directly in the
experimental width $\Gamma_{\mathit{exp}}$: Either the relevant
photoemission initial state does not depend on $k_\perp$, the electron
wave vector component perpendicular to the surface (as is the case for
2D surface bands) or the transition starts at symmetry points
$E_i(\vec{k})$ of the 3D bulk bands, where the band velocity $v =
\hbar^{-1}\mid\partial E/\partial k\mid$ is zero. To be more
quantitative the measured linewidth is given by \cite{smith93}
\begin{equation} 
\label{eq:Gamma} 
\Gamma_{\mathit{exp}} = \left(\Gamma_h + \frac{v_h}{v_e} \Gamma_e\right) 
\left(\left|1 - \frac{v_h}{v_e}\right|\right)^{-1} \   ,
\end{equation} 
where $\Gamma_h, \Gamma_e$ are the final state hole and electron
inverse lifetimes $\Gamma = \hbar/\tau$ and $v_h, v_e$ are the
corresponding band velocities. The detailed theory of photoemission,
using rigorous perturbation theory and incorporating $\Gamma_h$ and
$\Gamma_e$ in the relevant transition matrix elements, shows that an
isolated emission peak has a Lorentzian lineshape
\cite{matzdorf98,hufner95} provided initial and final state disperse
linearly across the peak and a quasiparticle description is
appropriate. In what follows we take normal emission spectra from
Ag(110) at different photon energies $\hbar\omega$, thereby tuning the
photoelectron transitions to $k$-space points around the X-point of
the $\Gamma$KX-direction parallel to the surface normal. At X, where
$v_h = 0$, equ.~(\ref{eq:Gamma}) results in $\Gamma_{\mathit{exp}} =
\Gamma_h$.  Additional mechanisms beyond hole decay may contribute to
the linewidth \cite{mcdougall95,matzdorf96,theilmann97,paggel99}. The
most important one is the hole-phonon interaction, which increases
$\Gamma_{\mathit{exp}}$ linearly with temperature $T$, except at very
low temperatures. By adequate extrapolation of our results to $T
\rightarrow 0$ we finally end up with an experimental upper limit for
the inverse hole lifetime $\hbar/\tau_h$.

\subsection{Data analysis} 
\begin{figure}[htbp]
  \begin{center}
   \includegraphics[width=12cm]{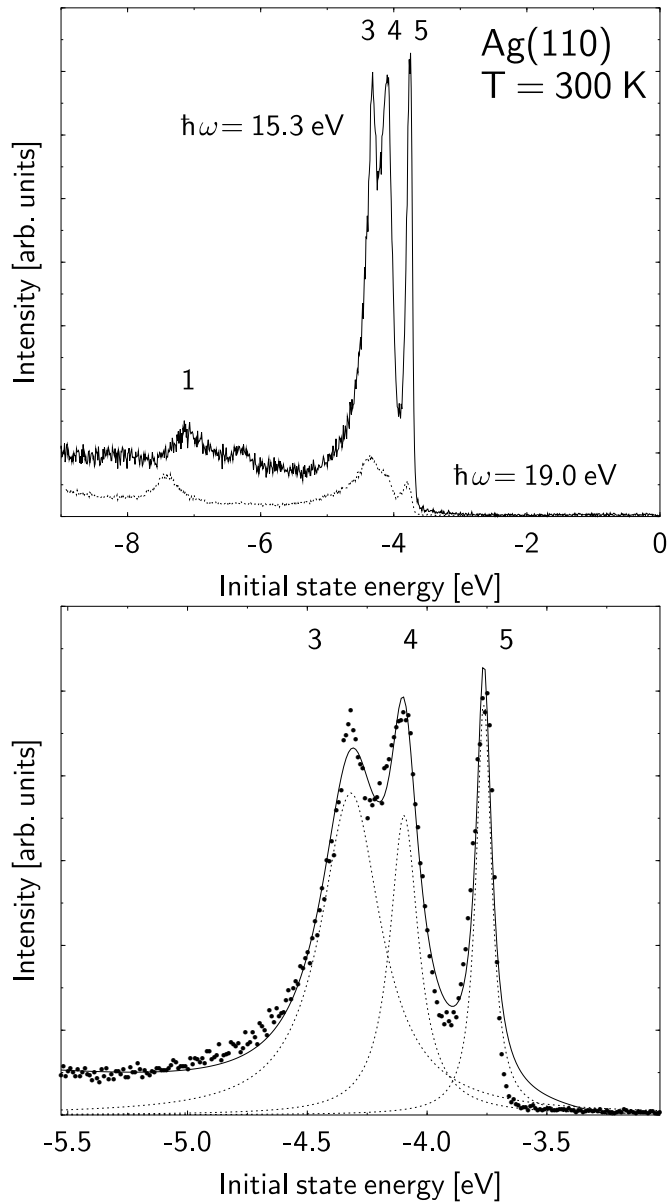} 
    \caption{Top: Normal-emission photoelectron spectra taken 
      at room temperature from Ag(110) with photons of energy
      $\hbar\omega = 15.3\,$eV (solid line) and $19.0\,$eV (dotted).  Note
      that both curves refer to the same intensity scale, obtained by
      normalization to the last-mirror current of the monochromator.
      Peak numbers refer to data collected in table 1. Bottom:
      Decomposition of peaks 3 to 5 using three Lorentzians and a
      linear background (not shown).  All energies given with respect
      to $E_F$.}
    \label{fig:edc308_46}
  \end{center}
\end{figure}

\begin{figure}[htbp]
  \begin{center}
    \includegraphics[width=12cm]{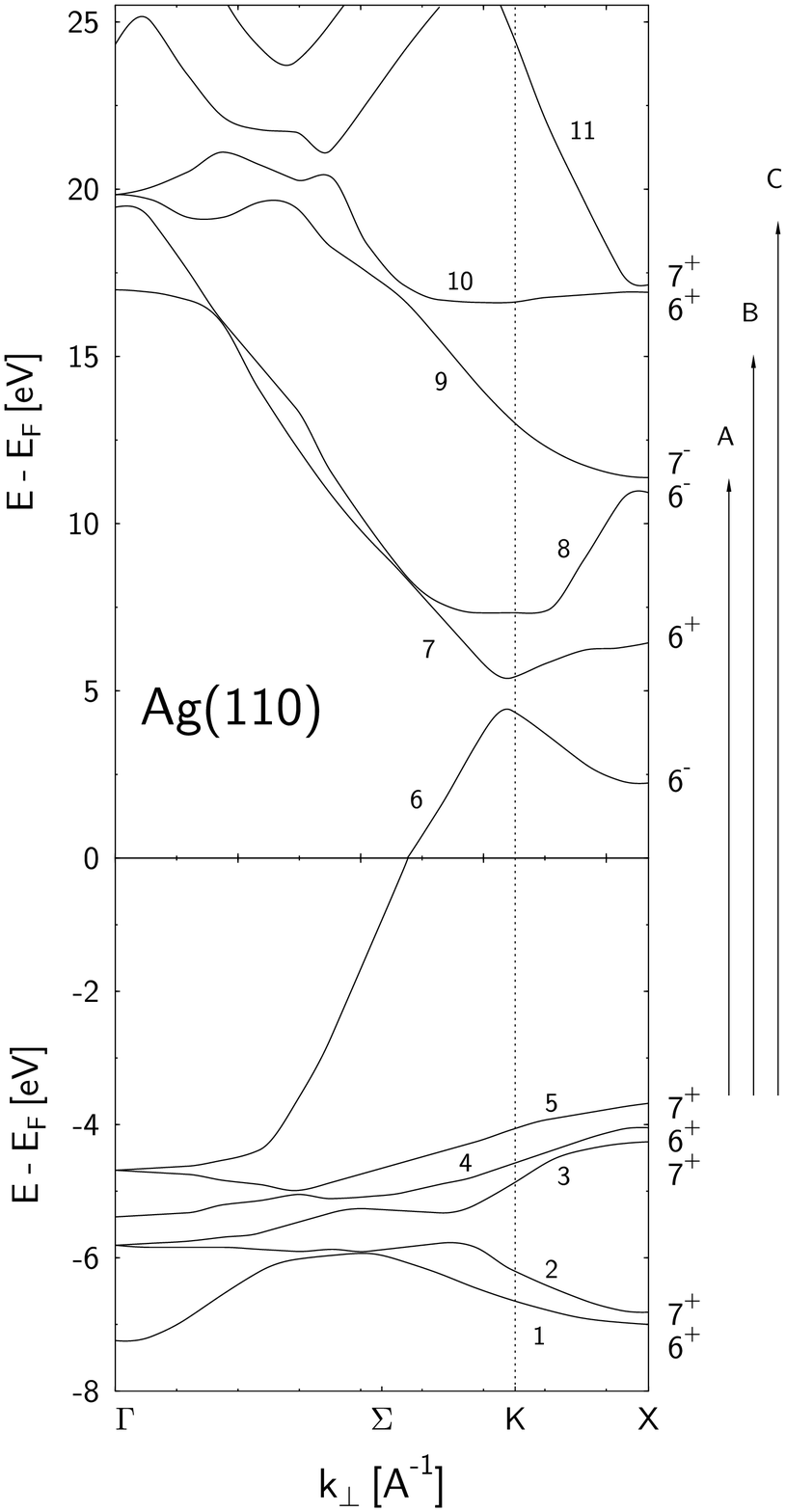}
    \caption{Bulk band structure of Ag calculated by Eckardt 
      et al.\ \cite{eckardt84} along the $\Gamma$KX-direction of the
      Brillouin zone. The bands have been labeled $1\ldots 11$
      counted from the lowest band below $E_F$ up to the highest
      energy above $E_F$. Note different energy scales above and below $E_F$.}
    \label{fig:ag110band}
  \end{center}
\end{figure}

We have collected normal emission spectra from Ag(110) for photon
energies between $\hbar\omega = 10.2\,$eV and $29.0\,$eV, both at room
temperature and $T = 22\,$K. The upper panel of Fig.\ 
\ref{fig:edc308_46} shows two spectra taken at $T = 300\,$K, with
$\hbar\omega = 15.3\,$eV (solid line) and $\hbar\omega = 19.0\,$eV
(dashed curve). These data were each normalized to the photon flux via
the current obtained from the last monochromator mirror.  Obviously
strong intensity changes occur with variation of $\hbar\omega$ which
will be discussed further below.  The bottom of Fig.\ 
\ref{fig:edc308_46} shows an enlarged part of the data taken at
$15.3\,$eV, together with three fit-curves based on a Lorentzian line
shape.  Three peaks $3\ldots5$ are clearly resolved which correspond
to emission from bulk $k$-space points near X$_{7^+},$ X$_{6^+}$ and
X$_{7^+}$, compare the band structure shown in Fig.\ 
\ref{fig:ag110band}. Peak 1 (upper panel) results from a direct
transition out of the bottom band between K and X.
\\

\begin{figure}[htbp]
  \begin{center}
    \includegraphics[width=12cm]{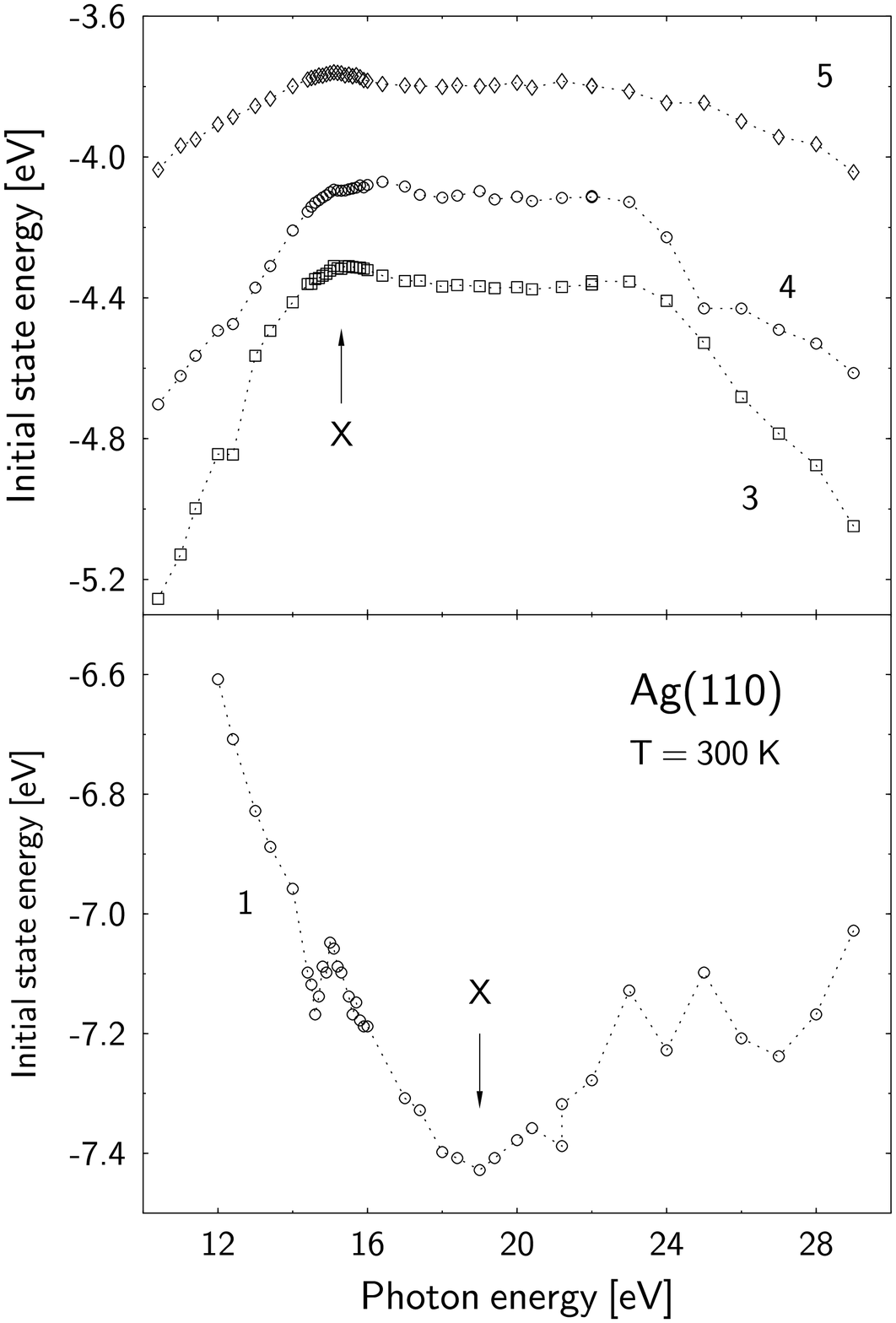}
    \caption{Initial state energies $E_i(\hbar\omega)$ observed 
      in normal emission from Ag(110) in their dependence on photon
      energy $\hbar\omega$. Numbers $1\ldots 5$ refer to initial state
      bands as labeled in Fig.\ \ref{fig:ag110band}.}
    \label{fig:disp}
  \end{center}
\end{figure}

From fits like those shown in Fig.\ \ref{fig:edc308_46} we obtain peak
positions as summarized in Fig.\ \ref{fig:disp}. As expected different
photon energies are necessary to induce transitions at the X-point:
The three upper $d$-bands around $E_i = -4\,$eV reach the critical
point at $\hbar\omega = 15.3\,$eV, easily identified by the smallest
binding energy $|E_i|$ in each band. In contrast the lowest $d$-band
is probed at the X-point with photons of $\hbar\omega = 19.0\,$eV,
corresponding to the greatest binding energy $|E_i|$. Our data resolve
only one peak from bands 1 and 2 near X. This is consistent with
earlier work: Wern et al.\ \cite{wern85} report a splitting of
$60\,$meV between X$_{6^+}$ and X$_{7^+}$ as deduced from data
collected at the Ag(100) surface.  Such a small splitting cannot be
resolved in our normal-emission spectra from Ag(110), because the
lifetime broadening exceeds the band splitting considerably, see below
for details. Our results for the various X-point energies are
collected in table 1.  Where comparable our results are in excellent
agreement with earlier band structure investigations of Ag
\cite{wern85,goldmann86}, see also table 7.2 in Ref.\ 
\onlinecite{hufner95}.
\\

\begin{figure}[htbp]
  \begin{center}
    \includegraphics[width=12cm]{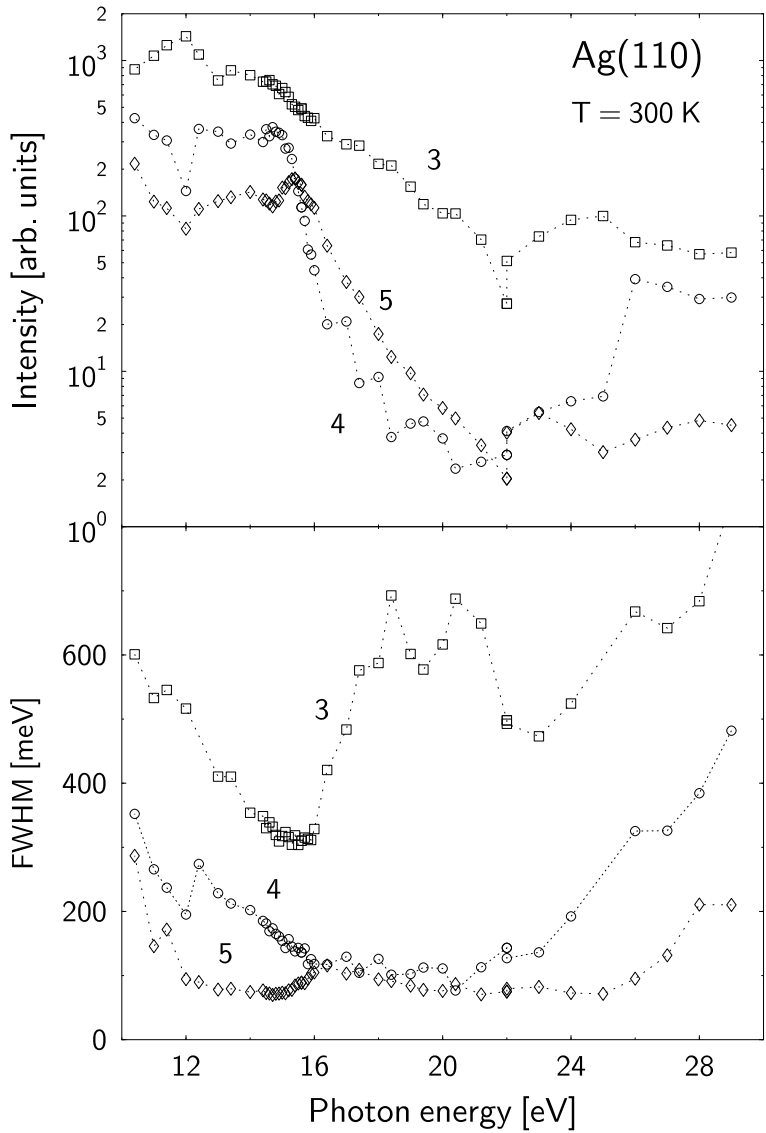}
    \caption{Top: Intensity variation of emission peaks 
      (labeled according to initial bands as numbered in Fig.\ 
      \ref{fig:ag110band}) as a function of $\hbar\omega$. Relative
      intensities obtained by normalization of count-rates to
      last-mirror current of the monochromator. Note that the ordinate
      scale is logarithmic.  Bottom: Full-width at half-maximum of the
      Lorentzians used to decompose the experimental spectra.}
    \label{fig:width_int}
  \end{center}
\end{figure}

From peak fits based on Lorentzians we can also determine the relative
intensities (area below peak) and the widths of the photoemission
lines in their dependence on the photon energy $\hbar\omega$. These
results are summarized in Fig.\ \ref{fig:disp}. Obviously the
intensities show a sudden drop at photon energies above $15\ldots
16\,$eV (top panel).  The interpretation results from closer
inspection of the band structure in Fig.\ \ref{fig:ag110band}: Below
$\hbar\omega = 15\,$eV the photoemission peaks $3 \ldots 5$ are due to
direct transitions from initial bands $3\ldots5$ to the final state
band 8, which carries essentially ``free-electron-like'' character,
i.e.\ in its wave-function the amplitude of the partial plane wave
exp$[i(\vec{k} + \vec{G}(110)) \cdot \vec{r}]$ is dominant. This is
the ``primary cone emission'' using Mahan's \cite{mahan70}
terminology, i.e.\ the final-state group velocity is directed along
the surface normal, with a concomitant high peak intensity. At
$\hbar\omega = 15.3\,$eV, however, the transition starting at the
X-point from band 5 (arrow A in Fig.\ \ref{fig:ag110band}) ends at the
lower edge of a gap opening between X$_{6^-}$ and X$_{7^+}$ (final
state energy $E_f = 10.9 \ldots 17.1\,$eV).  Therefore transitions
with $\hbar\omega = 19.0\,$eV from initial states with $E_i \approx
-4\,$eV near the X-point end within the gap (arrow B in Fig.\ 
\ref{fig:ag110band}) and no direct (vertical) transition is allowed.
Hence only ``surface emission'' (for more details see chapters 6.3.3
and 7.3.4 of Ref.\ \onlinecite{hufner95}) is possible, which conserves
$k_\parallel$ only.  The emission occurs via evanescent states located
energetically within the gap and one measures approximately a
one-dimensionally $k_\parallel$-resolved bulk density of initial
states.  Consequently not only intensities are affected (top panel of
Fig.\ \ref{fig:width_int}) but also the positions of peak $3\ldots 5$
(Fig.\ \ref{fig:disp}) remain essentially constant for transitions
into the gap region, i.e.\ at photon energies between about $16\,$eV
and $23\,$eV (arrow C in Fig.\ \ref{fig:ag110band}). At $\hbar\omega >
23\,$eV, direct transition are possible into band 11 and band
dispersion to larger $|E_i|$ is seen again in peaks $3\ldots 5$ (Fig.\ 
\ref{fig:disp}). A very similar behaviour, both with respect to photon
energy dependence of intensities and of experimentally observed
initial state energies, had already been observed for normal emission
from the Cu(110) surface \cite{dietz79}.  This is not surprising,
because Cu is isoelectronic and isostructural to Ag and exhibits a
similar band structure, with a final state gap between about 15 and
$20\,$eV above $E_F$ \cite{hufner95,dietz79}.
\\

\begin{figure}[htbp]
  \begin{center}
    \includegraphics[width=12cm]{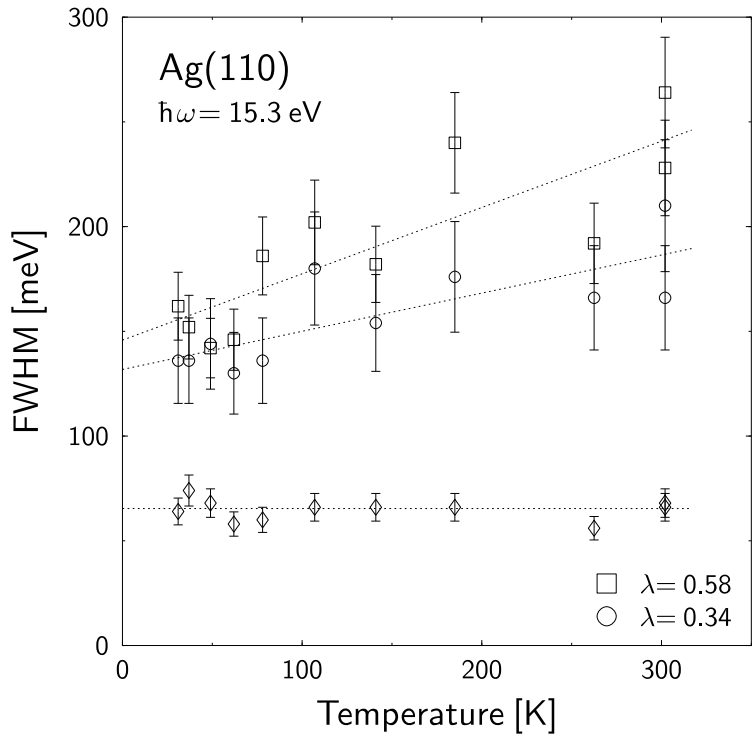}
    \caption{Experimental width (FWHM) of peaks $3\ldots 5$ of 
      Fig.\ \ref{fig:edc308_46} as a function of sample temperature.}
    \label{fig:agtemp}
  \end{center}
\end{figure}

Spectra with $\hbar\omega$ adjusted to initial states at the X-point
were also taken as a function of sample temperature. Two results are
of relevance in our context. Firstly the three initial state energies
$E_i$ of peaks $3\ldots 5$ shift only marginally to larger $|E_i|$
with increasing temperature: $\Delta E_i \le 20\,$meV between $T =
20\,$K and $310\,$K. This unexpectedly small effect results from a
compensation of two opposite trends: With increasing temperature the
lattice expands and the initial bands should move upwards in Fig.\ 
\ref{fig:ag110band}.  Simultaneously, however, the $d$-bands
experience reduced wave-function overlap, and this decreases the
overall $d$-band width.  Similar compensation has also been observed
at the X-point for Cu \cite{knapp79}, and the interpretation was based
on band-structure calculations for Cu at different lattice constant to
simulate thermal expansion \cite{knapp79}. The second temperature
effect is a linear increase in half-width of peaks 3 and 4, see Fig.\ 
\ref{fig:agtemp}. Its origin is the hole-phonon $(h-ph)$ interaction
\cite{mcdougall95,matzdorf96}. If we parameterize the increment in
width according to $\Gamma_{h-ph} = 2\pi \lambda k_BT$, we obtain
$\lambda = 0.58$ (peak 3) and $\lambda = 0.34$ (peak 4), respectively,
with an estimated error smaller than 15\%. Any phonon contribution to
the width of peak 5 is obviously below our experimental uncertainty of
$\pm 10\,$meV (corresponding to $\lambda < 0.15$).
\\

\begin{figure}[htbp]
  \begin{center}
    \includegraphics[width=12cm]{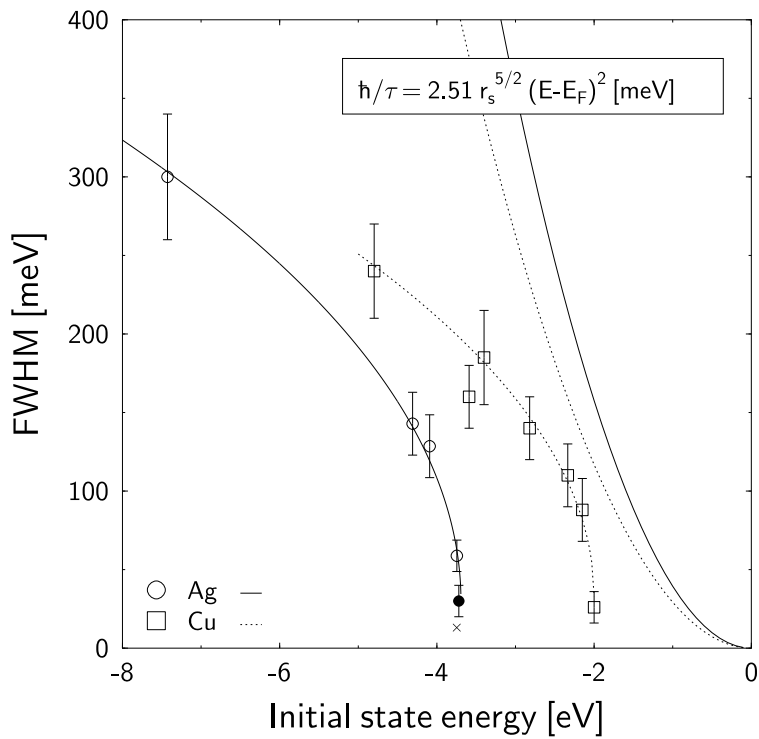}
    \caption{Experimental inverse $d$-hole lifetimes obtained 
      for Ag (open circles, this work) and for Cu (open squares, from
      Ref.\ \protect\onlinecite{gerlach10}) at the X-point of the bulk
      band structure, compare also table 1. The parabolas show
      calculations for Cu and Ag based on the free-electron gas (equ.\ 
      (\ref{eq:feg})). The filled circle for Ag results from a
      detailed line shape analysis (see text for details). The cross
      was obtained in Ref.\ \protect\onlinecite{luh00} from a $d$-like
      quantum-well state observed in normal emission from Ag layers on
      Fe(100).}
    \label{fig:agcuwidth}
  \end{center}
\end{figure}

Table 1 finally shows the experimental line-widths $\Gamma_h$ as well
as the corresponding hole lifetimes $\hbar/\Gamma_h$ after the
extrapolation to $T \rightarrow 0$ and a correction for the
experimental energy resolution. Additionally all data for $\Gamma_h$
are reproduced in Fig.\ \ref{fig:agcuwidth}, which will be discussed
further below. As mentioned already, peak 1 is a double-peak,
resulting from two transitions about $60\,$meV apart \cite{wern85}. We
have assumed equal intensity from both $d$-like initial states
X$_{6^+}$ and X$_{7^+}$ and subtracted $60\,$meV from the experimental
line-width.  This data is then shown in table 1 and in Fig.\ 
\ref{fig:agcuwidth}.
\\

\begin{table}[htbp] 
  \begin{center} 
    \begin{tabular}{c||crrr}
      \hline
      Peak & Symmetry & $E_i$ [eV] & $\Gamma_h$ [meV]& $\tau_h$ [fs] \\  \hline 
      1   & X$_{6^+},X_{7^+}$ & -7.43 & $300 \pm 40$ & 2 \\
      3   & X$_{7^+}$ & -4.31 & $143 \pm 20$ & 5 \\
      4   & X$_{6^+}$ & -4.07 & $129 \pm 20$ & 5 \\
      5   & X$_{7^+}$ & -3.76 & $58 \pm 10$  & 11 \\
      5   & X$_{7^+}$ & -3.70 & $30 \pm 10$  & 22 \\
      $q$ & $\overline{\Gamma}$ & -3.75 & 13 & 51 \\
      \hline
    \end{tabular} 
    \caption{Upper limits for $d$-hole lifetimes 
      $\tau_h$ due to electron-hole interaction at various symmetry
      points of the silver bulk bands. Peak $q$ is observed in Ref.\ 
      \protect\onlinecite{luh00} at $\hbar\omega = 12\,$eV in normal
      emission from Ag layers deposited on Fe(100) and results from a
      $d$-like quantum-well state.  Two results from our work are
      shown for peak 5: The first line is obtained from Fig.\ 
      \ref{fig:edc308_46}, the second line from the detailed peak
      shape analysis as described in the text.}
    \label{tab:lifetime} 
 \end{center} 
\end{table}  

\subsection{Line shape analysis} 
Closer inspection of Fig.\ \ref{fig:edc308_46} (bottom) shows that the
right wing of peak 5 is not adequately modeled by the shape of the
Lorentzian. This is not surprising, because there are at least two
questionable assumptions in this description: Firstly Lorentzian line
shape results only if both the initial and the final state bands
disperse linearly across the direct transition photoemission peak
\cite{smith93}. In our case this condition is clearly violated at the
X-point, compare Fig.\ \ref{fig:ag110band}.  Secondly $\Gamma_h$ and
$\Gamma_e$ were treated as constant within a spectral line. From a
theorists point of view $\Gamma_h$ and $\Gamma_e$ are the imaginary
parts of the self-energy which themselves depend on energy and
momentum. Especially at the upper $d$-band edge there might be a
significant variation of $\Gamma_h$ as suggested by our data in Fig.\ 
\ref{fig:agcuwidth}. Therefore we have performed a numerical line
shape calculation following the ideas outlined in more detail in
subsections 3.2 to 3.4 of Ref.\ \onlinecite{matzdorf98}.
\\

The basic strategy is as follows. The initial state bands
$E_i(\vec{k})$ are taken from the calculation of Ref.\ 
\onlinecite{eckardt84}.  In order to improve the agreement with the
experimental spectral line at $E_i = -3.76\,$eV they were rigidly
shifted downwards by $70\,$meV. The final state is modeled by a
free-electron-like band according to $E_f(k_\perp) = V_0 +
\hbar^2k_\perp^2/2m$, where $m$ is the free electron rest mass. We
have chosen $V_0 = -6.5\,$eV such that the X-point is crossed at
$\hbar\omega = 15.3\,$eV when considering peak 5.  We note, however,
that a variation of $V_0$ by $1\,$eV does not change the results within
our accuracy. The empirical bridging of the final state gap by a
free-electron-like band has been justified in many earlier studies,
see e.g.\ Ref.\ \onlinecite{nilsson79,grepstad82,slagsvold83} and many
references listed in Ref.\ \onlinecite{matzdorf98}. The finite escape
length of photoelectrons is equivalent to a damping, i.e.\ a non-zero
imaginary part of the lattice potential, which results in a complex
$k$-vector \cite{pendry74}. A sufficiently strong damping makes the
gap almost disappear and thereby gives rise to bands with an
approximated free-electron-like shape \cite{nilsson79}. This
assumption has been verified for Ag(110) in an earlier experimental
investigation \cite{goldmann86}. Since we only intend to model the
shape of the peaks we assume the photoemission matrix element, which
connects initial and final state, to be constant (i.e.\ independent of
$k_\perp$).
\\

The line shape $I(E_i)$ is then given for each contributing transition
by the convolution (equ.\ (20) of Ref.\ \onlinecite{matzdorf98})
\begin{equation} 
\label{eq:model}
I(E_i) \propto \int dk_\perp \, \mathcal{L}_h (E_f-E_i-\hbar\omega , \sigma_h) 
\mathcal{L}_e(k_\perp^0-k_\perp-G_\perp , \sigma_e).
\end{equation} 
In equ.~(\ref{eq:model}) $\mathcal{L}_h$ and $\mathcal{L}_e$ are
Lorentzian distributions, whereas $\sigma_h$ and $\sigma_e$ represent
the corresponding half-widths at half-maximum. They are treated as
parameters in order to reproduce the experimental line shape. After
performing the integration given by equ.~(\ref{eq:model}) the resulting
$I(E_i)$ is additionally convoluted with a Gaussian to account for the
experimental energy resolution before comparison with the measured
line shape.  To model $\sigma_h$ we have chosen
\begin{equation} 
\label{eq:sigma_h} 
\sigma_h(E_i) = 71 \left(|E_i|-3.70\right)^{1/2} + 15\  \textnormal{[meV]}
\end{equation} 
for the energy range $E_i < -3.70\,$eV. Above the upper $d$-band edge,
i.e.\ $E_i > -3.70\,$eV, we simply assume $\sigma_h = 15\,$meV. Our
choice of 2$\sigma_h = \Gamma_h$ is reproduced as the solid line drawn
through the FWHM results for Ag (open circles) in Fig.\ 
\ref{fig:agcuwidth} and assumes that $\Gamma_h$ does not change
significantly with $k_\perp$ in the vicinity of the X-point. This
parameterization, however, implies a drastic decrease of $\Gamma_h$ on
approaching the $d$-band edge. For the corresponding final states we
use
\begin{equation} 
\label{eq:sigma_e}
\sigma_e (E_f) = \frac{dk_\perp}{dE_f}\  \Gamma_e \ ,
\end{equation}
where $E_f$ is the free-electron-like parabola mentioned above. The
linear relation
\begin{equation} 
\label{eq:gamma_e} 
\Gamma_e = a (E_f-E_F)
\end{equation}
used in our calculation is an empirical average over many
photoemission results, see Ref.\ \onlinecite{goldmann91} for
experimental data and Ref.\ \onlinecite{echenique00} for some
theoretical background. Due to the dispersion of bands $3 \ldots 5$
around the X-point the particular choice of $a$ mainly influences the
left wing of peak 5.
\\

\begin{figure}[htbp]
  \begin{center}
    \includegraphics[width=12cm]{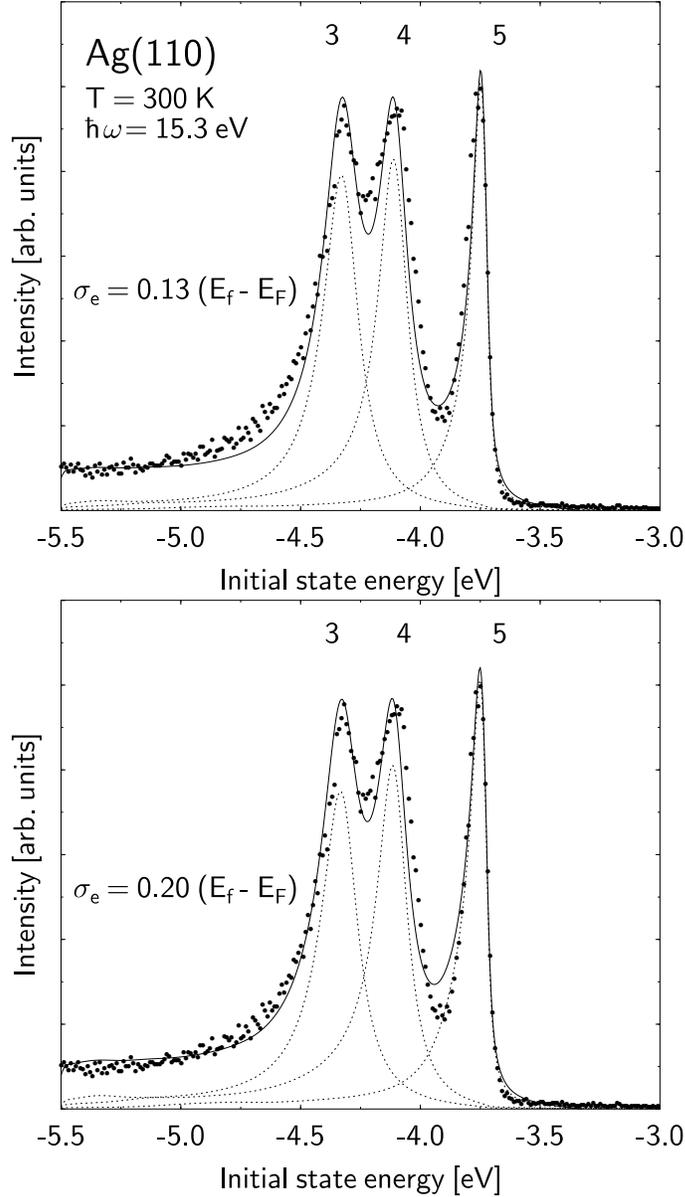} 
    \caption{Results from line shape analysis at the X-point
      using different broadening parameters for the photoelectron
      final state. See text for details.}
    \label{fig:upsint}
  \end{center}
\end{figure}

Typical results of the numerical integration following equations
(\ref{eq:model}) to (\ref{eq:gamma_e}) with $a=0.13$ and $a=0.20$ are
shown in Fig.\ \ref{fig:upsint}. Most insight is obtained by
considering the asymmetric peak 5 at the top of the $d$-bands.
Obviously the right wing of this peak is reproduced almost perfectly
in both cases. In particular the agreement of the data points and the
calculated peak (solid line) between $E_i = -3.5\,$eV and peak maximum
is very much improved by our choice of $\sigma_h(E_i)$ in this energy
range compared with the data shown in the bottom panel of Fig.\ 
\ref{fig:edc308_46}. A satisfying fit results with $a = 0.13$ (top
panel of Fig.\ \ref{fig:upsint}), whereas $a = 0.20$ overestimates the
asymmetry of peak 5 (bottom panel).  Similarly, calculations with $a <
0.13$ get the FWHM too small.  We have performed several calculations
varying also the other line-width parameters.  Our conclusions may be
summarized as follows.  At the upper $d$-band edge we observe a
significant decrease of $\Gamma_h$, which is consistent with the
measured line shape.  Consequently we get $\Gamma_h = 30 \pm 10\,$meV
at $E_i = -3.70\,$eV as upper limit.  The final state damping for
energies $E_f-E_F= 10 \ldots 12\,$eV is readily described with $a =
0.13 \pm 0.03$.  This number is in perfect agreement with our earlier
result obtained from an average over several metals \cite{goldmann91}.
In particular our analysis of peak 5 reveals that the asymmetric line
shape is due to a combination of hole-lifetime effects and integration
over a finite $k_{\perp}$-range around the X-point caused by the
electron-lifetime broadening of the final state.  Any variation of
$\Gamma_h(E_i)$ with $k_{\perp}$ in the vicinity of the X-point is
below our limits of detection.  Moreover, the line shape analysis of
peak 5 is consistent with an only small contribution to the linewidth
of less than $15\,$meV from scattering interactions with structural
defects.
\\

\section{Discussion} 
For many years the theory of quasiparticle lifetimes was based on
free-electron model calculations. Within this framework the inelastic
lifetime of hot electrons ($E > E_F$) as well as holes ($E < E_F$) is
approximately given \cite{quinn58} by
\begin{equation} 
\label{eq:feg} 
\tau = 263 \, r_s^{-5/2} (E - E_F)^{-2} \ \textnormal{fs eV$^2$,} 
\end{equation}
where the electron density parameter $r_s$ is defined by the relation
$1/n_0 = 4 \pi (r_s a_0^3)/3$ with the electron density $n_0$ and
$a_0$ being the Bohr radius. The strict validity of
equ.~(\ref{eq:feg}), however, is limited to energies close to $E_F$.
For comparison with experimental data we have plotted $\hbar / \tau$
according to equ.~(\ref{eq:feg}) for Ag ($r_s$ = 3.01) and Cu ($r_s$ =
2.67) in Fig.\ \ref{fig:agcuwidth}.\footnote{These values for $r_s$
  represent the electron density due to the $sp$-electrons only.} Also
included in this figure are the experimental results for Cu from Ref.\ 
\onlinecite{gerlach10}. The observed linewidths both for Cu and Ag are
clearly below the free-electron predictions.  As is evident, both
experimental data sets show a close similarity.  Particularly at
energies near the top of the $d$-bands the natural linewidth deviate
drastically from the quadratic energy dependence predicted by
equ.~(\ref{eq:feg}). This threshold-like behaviour has been first
observed for Cu in time-resolved two-photon photoemission
\cite{petek99} and was verified subsequently in one-photon
photoemission experiments \cite{matzdorf99}.
\\

We are not aware of any ab-initio calculations for lifetimes of
$d$-holes in Ag.  Nevertheless several first-principles calculations
\cite{echenique00,campillo00} based on the very details of the
electronic band structure have shown that the lifetimes generally
result from a delicate balance between density of states, localization
of electrons and dynamical screening. In this context Cu and Ag show
some remarkable differences: Due to the additional participation of
the Ag $4d$ states in the screening of the photohole longer lifetimes
might be expected. However, our results demonstrate that this effect
is compensated. Since the $d$-bands of Ag are located at $E_i \leq
-3.75\,$eV, i.e.\ almost $2\,$eV lower than in Cu, the enlarged phase
space of $sp$-electrons available for hole-scattering shortens the
lifetimes within the $d$-bands. This scattering should even be more
effective in Ag compared to Cu, because the Ag $4d$ electrons
experience a stronger delocalization.The measured relaxation times of
hot electrons in Ag with $E-E_F < 3.7\,$eV essentially follow the
energy dependence predicted by the free-electron model
\cite{aeschlimann96}.  Especially the characteristic long lifetimes of
noble metals, i.e.\ $50\,$fs at $1\,$eV above the Fermi level, are observed.
Due to the large threshold for creation of $d$-holes these electrons
do not show up in time-resolved photoemission.
\\

The hot electron and hole-dynamics has recently been analyzed for Cu
and Au by means of first-principles many-body calculations taking the
$d$-bands explicitly into account \cite{campillosubmitted}. In
agreement with the experiment these results indicate that $d$-holes in
Cu and Au exhibit larger inelastic lifetimes than excited
$sp$-electrons at corresponding distance from $E_F$.  This enhancement
is in particular strong at the upper $d$-band edge, i.e.\ the
threshold for creation of $d$-holes. The underlying physics is as
follows \cite{campillosubmitted}: While the density of states
available for $d$-hole decay is larger than that for the decay of
excited electrons (this should shorten the lifetimes) the scattering
cross-section of $d$- and $sp$-states below $E_F$ is very small at the
top of the $d$-bands, and this considerably decreases the decay rates.
With increasing distance to $E_F$ ($E_i < -2\,$eV in Cu) the phase
space for the hole to decay increases and, simultaneously, the
cross-section for hole-scattering within the $d$-bands gets larger
\cite{campillosubmitted}. In consequence the linewidths increase below
the upper $d$-band edge, in agreement with the experimental data shown
in Fig.\ \ref{fig:agcuwidth}.
\\

As mentioned already, calculations are available for Cu only. At
present a quantitative comparison of experiment and theory is still
difficult because of two critical discrepancies: Firstly these
calculations get the $d$-bands rigidly shifted up by $0.5\,$eV as
compared to experiment.  Secondly the calculated decay rates usually
represent an average over all wave vectors and electron bands
referring to the same initial state energy.  There is one exception,
however. At the top of the Cu $d$-bands, at the X$_5$ point
(calculated at $E_i = -1.5\,$eV) and with the $k$-vector along the
$\Gamma$X-direction, theory obtains a $d$-hole lifetime of $99\,$fs
\cite{campillosubmitted}. This has to be compared to the $k$-space
averaged value of $72\,$fs at the same energy below $E_F$
\cite{campillosubmitted}. The lifetime of $\tau = 99\,$fs corresponds
to a predicted photoemission linewidth $\Gamma_h = 7\,$meV at X$_5$,
which is clearly compatible with the experimental upper limit of Fig.\ 
\ref{fig:agcuwidth}, i.e.\ $\Gamma_h \le 25\,$meV. We also mention in
this context that our result for the upper $d$-band edge at of Ag is
$\Gamma_h \le 30\,$meV, and that the quantum well state observed in Ag
layers on Fe(100) may be modeled using a lifetime width of $\Gamma_h =
13\,$meV \cite{luh00}, see the cross in Fig.\ \ref{fig:agcuwidth}. The
latter numbers yield $\tau_h \geq 22\,$fs and $\tau_h \geq 51\,$fs,
which data appear reasonable in the light of the Cu calculations.
\\

In summary we have presented a detailed experimental investigation of
inverse $d$-hole lifetimes at the X-point of Ag. The data show a
threshold-like behaviour at the upper edge of the $d$-bands, with
unexpectedly long lifetimes $\tau_h \geq 22\,$fs at $3.70\,$eV below
$E_F$. For deeper lying bands, the experimental $\tau$-values get much
smaller, down to $2\,$fs at $7.43\,$eV below $E_F$. This trend is similar to
what has been observed for copper and demonstrates convincingly that
free-electron-like models are not adequate to describe $d$-hole decay
in noble metals. As shown by calculations for Cu the lifetime
enhancement at the upper $d$-band edge results from very small
scattering cross-section of $d$- and $sp$-states. With increasing
distance to $E_F$, the rapidly increasing density of states and
intraband scattering enhance the decay rate and thus shorten the
$d$-hole inelastic lifetimes considerably. We hope that our result
also stimulates more detailed theoretical work, because quantitative
understanding of hole decay is an important ingredient in any attempt
to tailor excited electron dynamics.

\section*{Acknowledgement} 
Our work at Bessy is supported by the German Bundesministerium f\"ur
Bildung und Forschung (BMBF). Preparation procedures to obtain
low-defect sample surfaces were carefully investigated before in our
home laboratory. This work is continuously supported by the Deutsche
Forschungsgemeinschaft (DFG). We thank W.\ Braun and the Bessy staff
for their important beamline support.

\end{document}